\documentclass[twocolumn,showpacs,preprintnumbers,amsmath,amssymb]{revtex4}

\newcommand{\bib}{\bibitem}
\newcommand{\bea}{\begin{eqnarray}}
\newcommand{\eea}{\end{eqnarray}}
\newcommand{\beq}{\begin{equation}}
\newcommand{\eeq}{\end{equation}}
\newcommand{\non}{\nonumber}
\newcommand{\da}{\dagger}
\newcommand{\al}{\alpha}

\newcommand{\ga}{\gamma}

\newcommand{\la}{\lambda}
\newcommand{\om}{\omega}
\newcommand{\si}{\sigma}
\newcommand{\pa}{\partial}
\usepackage{psfrag,graphicx,epsfig} 
\usepackage{bm} 

\begin{document}

\title{AC conductivity of a quantum Hall line junction}
\author{Amit Agarwal and Diptiman Sen}
\affiliation{Center for High Energy Physics, Indian Institute of Science, 
Bangalore 560 012, India}
\date{\today}

\begin{abstract}
We present a microscopic model for calculating the AC conductivity of a 
finite length line junction made up of two counter or co-propagating single 
mode quantum Hall edges with possibly different filling fractions. The effect
of density-density interactions and a local tunneling conductance ($\si$) 
between the two edges is considered. Assuming that $\si$ is independent 
of the frequency $\om$, we derive expressions for the AC 
conductivity as a function of $\om$, the length of the line 
junction and other parameters of the system. We reproduce the results of 
\prb{\bf 78}, 085430 (2008) in the DC limit ($\om \to 0$), and 
generalize those results for an interacting system. As a function of $\om$, 
the AC conductivity shows significant oscillations if $\si$ is small; the 
oscillations become less prominent as $\si$ increases. A renormalization 
group analysis shows that the system may be in a metallic or an insulating 
phase depending on the strength of the interactions. We discuss the 
experimental implications of this for the behavior of the AC conductivity 
at low temperatures.
\end{abstract}

\maketitle

\section{Introduction}

A line junction (LJ) \cite{renn,oreg,kane1,mitra,kollar,kim,zulicke,papa,das}
separating two edges of fractional quantum Hall (QH) states
allows the realization of one-dimensional systems of interacting electrons
for which the Luttinger parameter can be tuned \cite{gogolin,giamarchi1,rao}.
A LJ is formed by using a gate voltage to create a narrow barrier which
divides a fractional QH state such that there are two chiral edges flowing in 
opposite directions (counter propagating) on the two sides of the barrier 
\cite{kang,yang1,yang2,roddaro1,roddaro2}. For a QH system corresponding to 
a filling fraction which is the inverse of an odd integer such as $1, 3, 5, 
\cdots$, the edge consists of a single mode which can be described by a 
chiral bosonic theory \cite{wen1}. In a system with two QH states separated 
by a line junction, the edges on the two sides of the LJ generally 
interact with each other through a short-range density-density interaction 
(screened Coulomb repulsion); such an interaction can be treated exactly in 
the bosonic language. The physical separation between the two edges and, 
therefore, the strength of the interaction can be controlled by a gate 
voltage. In general, a LJ also allows tunneling between the two edges; if the
LJ is disordered, the tunneling amplitude is taken to be a random variable.
The tunneling amplitude is also dependent on the separation between the edges.

Recently, QH systems with a sharp bend of $90^0$ have been 
fabricated \cite{grayson1,grayson2}. An application of an appropriately 
tilted magnetic field in such a system can produce QH states on the two 
faces which have different filling fractions $\nu_1$ and $\nu_2$, since
the filling fractions are governed by the components of the magnetic field
perpendicular to the faces. The two perpendicular components can even have 
opposite signs if the magnetic field is sufficiently tilted. Depending on
whether $\nu_1$ and $\nu_2$ have the same sign or opposite signs, the edge 
states on the two sides of the line separating the two QH states may 
propagate in opposite directions or in the same direction; we call these 
counter or co-propagating edges respectively. A QH system with a bend
therefore provides a new kind of LJ in which the filling fractions can be 
different on the two sides of the LJ, and the two edges can be co-propagating.

In an earlier paper \cite{sen}, we developed a microscopic model for the
direct current (DC) conductivity of a finite length LJ with either counter
or co-propagating edges. The conductivity is expressed by a current
splitting matrix $S_{dc}$ which depends on the filling fractions $\nu_1$ and 
$\nu_2$, the choice of current splitting matrices which provide boundary 
conditions for the bosonic fields at the two ends of the LJ, the 
tunneling conductance per unit length $\si$, and the length 
$L$ of the LJ. The Coulomb interaction between the edges was ignored in the 
calculation of the DC conductivity, but the effect of the interaction was 
then taken into account to study the renormalization group flow of the 
tunneling conductance and therefore the conductivity.

In this paper, we will generalize the results of Ref. \cite{sen} to
find the alternating current (AC) conductivity along a LJ; the inter-edge
interactions will be taken into account in this calculation. In the limits 
in which the AC frequency $\om$ goes to zero, we will recover the 
results obtained in Ref. \cite{sen}. 

The paper is organized as follows. In Sec. II, we introduce a microscopic 
model for the LJ and discuss the general form of the current splitting matrix 
$S$ for both DC and AC. We obtain the condition for zero power dissipation. 
We also discuss the possible boundary conditions which can be imposed
at the ends of the line junction; if we require that the commutation
relations of the incoming and outgoing bosonic fields be preserved,
we find that the boundary conditions must take one of two forms, which are
described by matrices $S_0$ and $S_1$. In Sec. III, we introduce short-range 
interactions (whose strength is given by a parameter $\lambda$) and a
local tunneling conductance (denoted by $\si$) between
the edges of the line junction. We then discuss the case of counter 
propagating edges and present the frequency dependent matrix $S_{ac}$ 
for some simple choices of the filling fractions and velocities of the edge 
modes. In Sec. IV, we discuss the case of co-propagating edges and present 
the matrix $S_{ac}$, again for some simple cases. We present some plots of 
the elements of $S_{ac}$ as functions of the AC frequency $\om$. In Sec. V, 
we discuss the implications of a renormalization group analysis for the low 
temperature behavior of $S_{ac}$, and how this may be checked experimentally. 
Section VI summarizes our results and discusses possible extensions of our 
work. In the Appendix we present the details of the calculations for the 
general case of arbitrary filling fractions and velocities of the edge modes.

\section{Model for the line junction}

We consider a LJ with two different QH liquids on the two sides. The edges 
of the QH liquids on the two sides of the LJ are assumed to be spatially 
close to each other; hence there are density-density interactions
between the two edges, and electrons can also tunnel between the edges. 
For simplicity, we will assume that the interaction strength and the 
tunneling conductance have the same magnitude at all points along the LJ.
We will also assume that the incoming and outgoing fields connect 
continuously to the corresponding fields at each end of the LJ.

Consider the counter propagating (co-propagating) LJ systems shown in Figs. 
1 (a) and (b) respectively. 
The currents (voltages) in the two incoming edges are denoted as $I_1$ ($V_1$)
and $I_2$ ($V_2$), and in the two outgoing edges as $I_3$ ($V_3$) and $I_4$ 
($V_4$). Here edges $1$ and $3$ correspond to a QH system with filling 
fraction $\nu_1$, while edges $2$ and $4$ correspond to a system with filling 
fraction $\nu_2$. We also assume that the QH edge modes are locally 
equilibrated; discussions of equilibration at zero frequency have been 
presented in Refs. \cite{kane2,buttiker}. Namely, at each point $x$, which may
lie either on one of the outer edges 1-4 or inside the line junction (where 
$x$ goes from 0 to $L$), we assume, for small bias, that
\beq I_{i} (x,t) ~=~ \frac{e^2}{h} ~\nu_i ~V_{i} (x,t). \label{iv} \eeq
In the linear response regime (when the applied voltage bias is small), 
we expect the outgoing currents to be 
related to the incoming ones in a linear way. Let us denote the alternating 
current on edge $i$ by $I_i = \al_i e^{i(k_i x - \om t)} + c.c.$, where 
$\al_i$ is a complex number in general. The numbers $\al_i$ 
are related by a current splitting matrix $S_{ac} (\om)$ as 
\bea \left( \begin{array}{c} \al_3 \\
\al_4 \end{array} \right) &=& S_{ac} ~\left( \begin{array}{c} \al_1 \\
\al_2 \end{array} \right), \non \\
{\rm where}~~~~ S_{ac} &=& \left( \begin{array}{cc} r(\om) & \bar{t}(\om) \\
t(\om) & \bar{r}(\om) \end{array} \right). \label{sac} \eea

When all the edge states are in equilibrium, the power dissipated is given by 
the difference of the incoming and outgoing powers, namely,
\bea \label{P1}
P &=& \frac{1}{2} ~[I_1 V_1 ~+~ I_2 V_2 ~-~ I_3 V_3 ~-~ I_4 V_4]. \eea
If there is no power dissipation in the system (we will see below that this is
true if the tunneling conductance is zero all along the LJ), then the average 
over one oscillation cycle of the incoming energy must be equal to the 
outgoing energy. This imposes the following condition
\beq S_{ac}^\da ~\left( \begin{array}{cc} 1/\nu_1 & 0 \\
0 & 1/\nu_2 \end{array} \right) ~S_{ac} ~=~ \left( \begin{array}{cc} 
1/\nu_1 & 0 \\
0 & 1/\nu_2 \end{array} \right), \label{cons1} \eeq
or, more explicitly,
\bea \frac{|r(\om)|^2}{\nu_1} ~+~ \frac{|t(\om)|^2}{\nu_2} &=& 
\frac{1}{\nu_1}, \non \\
\frac{|\bar{r}(\om)|^2}{\nu_2} ~+~ \frac{|\bar{t}(\om)|^2}{\nu_1} &=& 
\frac{1}{\nu_2}, \non \\
{\rm and} ~~~~\frac{r^* (\om) \bar{t} (\om)}{\nu_1} ~+~ \frac{t^* (\om) 
\bar{r}(\om)}{\nu_2} &=& 0. \eea

When the incoming currents are DC in nature, the currents satisfy a linear 
relation and are related by a current splitting matrix $S_{dc}$. This is a 
real matrix which can be characterized by a single parameter $\ga$ (called
the scattering coefficient); it has the general form \cite{wen2,halperin,sen}
\beq \left( \begin{array}{c} I_3 \\ I_4 \end{array} \right) ~=~ 
\left( \begin{array}{cc} 1 - \frac{2 \ga \nu_2}{\nu_1+\nu_2} &
\frac{2 \ga \nu_1}{\nu_1+\nu_2} \\ \frac{2 \ga \nu_2}{\nu_1+\nu_2} & 
1-\frac{2 \ga \nu_1}{\nu_1+\nu_2} \end{array}
\right) \left( \begin{array}{c} I_1 \\
I_2 \end{array} \right). \label{sdc} \eeq
The power dissipated is the difference of the incoming and outgoing energy 
flux (\ref{P1}), and it is given by
\beq P ~=~ \frac{e^2}{h} ~\frac{2\nu_1 \nu_2}{\nu_1 + \nu_2} ~\ga (1 - \ga)~ 
(V_1 - V_2)^2. \eeq
The condition that $P \ge 0$ requires that $0 \le \ga \le 1$. No power is 
dissipated if $\ga = 0$ or $1$, and maximum power dissipation occurs when
$\ga = 1/2$.

\begin{figure}[t] \begin{center}
\includegraphics[width=1.0 \linewidth]{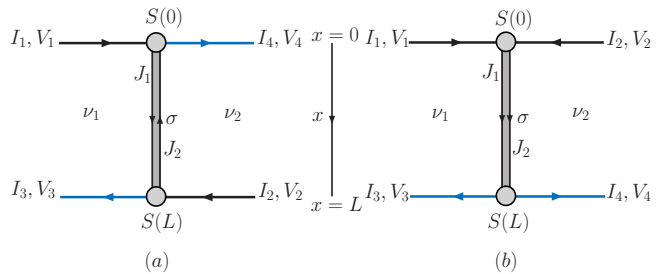} \end{center}
\caption{(Color online) Schematic picture of a line junction with (a) counter 
propagating and (b) co-propagating edges, with two incoming and two outgoing 
edges.} \end{figure}

The end points of the LJ shown in Fig. 1 lie at $x=0$ and $L$. At each of 
these ends, we have two incoming edges and two outgoing edges; two of these 
correspond to the outer edges marked $I_1$, $I_2$, $I_3$, and $I_4$, while 
the other two edges are internal to the LJ and are marked $J_1$ and $J_2$.
An important ingredient of the model for such a system is the
boundary condition which should be imposed at the end points. In Ref. 
\cite{sen}, it was shown that there are two possible boundary conditions
which can be imposed at each end; both of these allow us to smoothly connect
the bosonic fields which may be used to calculate the currents in the system. 
The two possible boundary conditions correspond to using one of two matrices
$S_0$ or $S_1$ to related the incoming and outgoing modes at each end, where
\bea S_0 &=& \left( \begin{array}{cc} 1 & 0 \\
0 & 1 \end{array} \right), \non \\
{\rm and} ~~~ S_1 &=& \frac{1}{\nu_1 + \nu_2} ~\left( \begin{array}{cc}
\nu_1 - \nu_2 & 2 \nu_1 \\
2 \nu_2 & \nu_2 - \nu_1 \end{array} \right). \label{s01} \eea
In this paper, we will only consider the boundary condition corresponding to
the matrix $S_0$; this is physically the more plausible boundary condition 
since it just connects the incoming edge to the outgoing one for each 
QH liquid separately.

\section{Counter propagating case}

We will now present a microscopic model of a LJ for the case of counter 
propagating edges. For simple filling fractions $\nu_i$ given by the inverse 
of an odd integer, each edge is associated with a single chiral boson mode. 
For the counter propagating case, shown in Fig. 1 (a), the mode on one edge 
propagates from $x=0$ to $x=L$, while the mode on the other edge propagates 
in the opposite direction; let us call the corresponding bosonic fields as 
$\phi_1$ (right mover) and $\phi_2$ (left mover) respectively. In the absence 
of density-density interactions between these edges, the Lagrangian is given by 
\bea {\cal L} &=& \frac{1}{4\pi \nu_1} ~\int_0^L dx ~\pa_x \phi_1~ (- \pa_t -
v_1 \pa_x) ~\phi_1 \non \\
& & + ~\frac{1}{4\pi \nu_2} ~\int_0^L dx ~\pa_x \phi_2 ~(\pa_t - v_2 \pa_x) ~
\phi_2 , \label{lag1} \eea
where $v_i$ denotes the velocity of mode $i$. The density and current fields 
are defined as
\bea \begin{array}{cccc} 
\rho_1 ~=& \pa_x \phi_1/(2\pi), ~~& J_1 ~=& ~ -\pa_t \phi_1/(2\pi), \\
\rho_2 ~=& - ~\pa_x \phi_2/(2\pi), ~~& J_2 ~=& \pa_t \phi_2/(2\pi). 
\end{array} \eea
For a short-range density-density interaction between the two edges, the term 
in the Lagrangian is of the form 
\bea {\cal L}_{int} &=& \frac{\la}{4\pi \sqrt{\nu_1 \nu_2}} ~\int_0^L 
dx ~\pa_x \phi_1~ \pa_x \phi_2, \label{lag2} \eea
where $\la$ is the interaction strength (positive for repulsive interactions) 
with the dimensions of velocity. 

The equations of motion for the Lagrangian given in Eqs. (\ref{lag1}) and 
(\ref{lag2}), written in terms of the density and the current fields, are
\bea J_1 ~-~ v_1 \rho_1 ~-~ \frac{\lambda \nu_1}{2 \sqrt{\nu_1 \nu_2}}~
\rho_2 ~=~ 0, \non \\
J_2 ~+~ v_2 \rho_2 ~+~ \frac{\lambda \nu_2}{2 \sqrt{\nu_1 \nu_2}}~ \rho_1 ~=~
0. \label{eqm1} \eea

A model of tunneling at zero frequency between different edges or point 
contacts in a QH system has been developed in Ref. \cite{kane2}. By adding
a time derivative term in their expressions, we can model tunneling at finite
frequencies between the two edges along the LJ using the following equations
\bea \pa_t \rho_1 ~+~ \pa_x J_1 &=& \frac{\si h}{e^2}\left(\frac{J_2}{\nu_2}
-\frac{J_1}{\nu_1} \right), \non \\ 
\pa_t \rho_2 ~+~ \pa_x J_2 &=& \frac{\si h}{e^2}\left(\frac{J_2}{\nu_2}
-\frac{J_1}{\nu_1} \right), \label{eqm2} \eea
where $\si$ is the conductance per unit length across the LJ. Physically, Eq.
(\ref{eqm2}) is the continuity equations for the edge states with a source 
term \cite{kane2}, the source term being the current tunneling into the system
because of the voltage difference between the corresponding points on the line
junction, $I_{source}=\si (V_2-V_1)=(\si h/e^2) (J_2 /\nu_2 - J_1 /\nu_1)$.
We will assume $\si$ to be constant along the LJ. Unlike Eq. (\ref{eqm1}), 
the model of tunneling given in Eq. (\ref{eqm2}) cannot be derived from any 
Lagrangian since it is non-unitary, and a non-zero value of $\si$ implies 
that there is dissipation in the system.

For the DC case, in a non-interacting system, the current splitting matrix
is given by Eq. (\ref{sdc}), and $\ga$ is given by \cite{sen}
\beq \label{tdc1} 
\ga = \frac{\nu_1+\nu_2}{2} \frac{1-e^{-L/l_c}}{\nu_2 -\nu_1e^{-L/l_c}} ~~~
{\rm and} ~~~ l_c^{-1} = \frac{\si h}{e^2} \left( \frac{1}{\nu_1} -
\frac{1}{\nu_2} \right) \eeq
when $\nu_1 \neq \nu_2$. For the special case \cite{kane1} of $\nu_1 = \nu_2 
= \nu$, we obtain $\ga = [1 + \nu e^2/(\si L h)]^{-1}$.

Now we solve the problem for the general case with interactions and for an
arbitrary value of $\om$. We can combine Eqs. (\ref{eqm1}) and (\ref{eqm2}) 
to obtain
\bea \label{eqm3}
\left(\pa_t+v_1 \pa_x + \frac{\al}{\nu_1}\right)J_1 &+& \left( \frac{\la 
\nu_1}{2 \sqrt{\nu_1\nu_2}} \pa_x - \frac{\al}{\nu_2} \right) J_2=0, \non \\
\left(\pa_t-v_2 \pa_x + \frac{\beta}{\nu_2}\right)J_2 &-& \left( \frac{\la 
\nu_2}{2 \sqrt{\nu_1\nu_2}} \pa_x + \frac{\beta}{\nu_1}\right)J_1=0, \non \\
\eea
where 
\bea \al &=& \frac{\si h}{e^2} ~(v_1+\frac{\la \nu_1}{2 \sqrt{\nu_1\nu_2}}),
\non \\
\beta &=& \frac{\si h}{e^2} ~(v_2+\frac{\la \nu_2}{2 \sqrt{\nu_1\nu_2}}). \eea
Solving these equations with appropriate boundary conditions gives us the 
current splitting matrix $S_{ac}$. We will work with the boundary condition 
that connects the fields along the LJ continuously to the corresponding
incoming and outgoing fields, i.e., the incoming field $I_{1/2}=J_{1/2}(0)$ 
at $x=0$ and $J_{1/2}(L) = I_{3/4}$ at $x=L$. The most general case will 
have $\nu_1 \neq \nu_2 $ and $v_1 \neq v_2$. We solve Eq. (\ref{eqm3}) in its
most general form in the Appendix and present the matrix $S_{ac}$.
In this section we present results for some relatively simple cases.

For the case of LJ with interactions but no tunneling ($\si=0$), same filling
fraction ($\nu_1=\nu_2=\nu$) and same velocity ($v_1=v_2=v$), we find that
\bea t(\om) &=& \bar{t}(\om) ~=~ - ~\frac{\lambda \sin (k L)}{2 \left[ 
i \tilde{v} \cos (k L) + v \sin (k L) \right]}, \non \\
r(\om) &=& \bar{r}(\om) ~=~ \frac{i \tilde{v}}{i\tilde{v} \cos (k L) + v 
\sin (k L)}, \label{trom} \eea
where $k=\om/{\tilde v}$ and ${\tilde v}=\sqrt{v^2-\lambda^2/4}$.
Note that there is no dissipation in this case, and the AC current 
splitting matrix satisfies Eq. (\ref{cons1}). We also note that this 
solution is consistent with Eq. (11) of Ref. \cite{oreg} in the limit 
of $\la << v$, (our $r(\om)$ corresponds to their $t(\om)$). Similar 
expressions have also appeared in Refs. \cite{safi2,safi3}. Also note that 
in the DC limit ($\om \to 0$), $|r(\om)|=1$. This implies that conductance 
across the LJ, $G=\nu e^2/h$. This is expected and is consistent with 
Ref. \cite{maslov,ponomarenko,safi1,safi4}.

For the case of same $\nu$'s and same velocities but with the tunneling $\si$
switched on, we have
\bea t(\om) =& \frac{2 \left(k^2 v^2 \nu ^2+(\al -i \nu \om )^2
\right)}{k \nu (2 v \al +\lambda (\al -i \nu \om )) \cot (k L) + 
\left(-k^2 v \lambda \nu^2 +2 \al (\al -i \nu \om ) \right)}, \non \\ 
r(\om) =& \frac{1}{\cos (k L)+\frac{\left(-k^2 v \lambda \nu^2 + 2 \al
(\al -i \nu \om) \right) \sin (k L)}{k \nu ( 2 v \al +\lambda
(\al -i \nu \om ))}}, \eea
where $\al = (\si h /e^2) \left (v+\lambda/2 \right)$, and 
\beq k ~=~ \frac{\om}{\sqrt{v^2-\la^2/4}} ~\left(1+\frac{2 i \si h}
{\nu e^2 \om}(v+\lambda/2)\right)^{1/2}. \label{kom} \eeq 

The expression for the most general case is given in the Appendix in Eq.
(\ref{KA3}) and (\ref{TRA5}). In the DC limit $\om \to 0$, Eq. (\ref{KA3})
gives $k_1 \to il_c^{-1} = i(\si h /e^2) (\nu_1^{-1} - \nu_2^{-1})$ and 
$k_2 \to 0$, while Eq. (\ref{TRA5}) gives 
\beq \label{r1}
r(\om \to 0) ~=~ \frac{\nu_1 - \nu_2}{\nu_1 - \nu_2 ~ e^{L/l_c}}. \eeq
Comparing with the expression in Eq. (\ref{sdc}), we get the same value of 
$\ga$ as in Eq.(\ref{tdc1}). 

In Fig. 2, we show the absolute values of the various reflection and 
transmission amplitudes as functions of the frequency $\om$ (in units of 
$v_1/L$ which has been set equal to unity) for various choices of the filling 
fractions $\nu_i$, velocities $v_i$, length $L$, interaction $\la$ (in units 
of $v_1 = 1$), and tunneling conductance per unit length $\si$ (in units of 
$e^2/(hL)$). Figs. 2 (a) and (b) show the cases of
$\si = 0$ (zero tunneling) for equal filling fractions and different filling 
fractions respectively; in Fig. 2 (a), $|r| = |\bar r|$ and $|t| = |\bar t|$ 
by symmetry. In both figures, we see prominent oscillations as a function of 
$\om$. This is clear from Eq. (\ref{trom}) where we see that $k$ is real and 
the different amplitudes oscillate with a wavelength $2\pi /k$. In the $\om 
\to 0$ limit, $|r| = |\bar{r}| = 1$ for both cases, as is evident from Eq. 
(\ref{r1}). Both these cases are dissipationless and the amplitudes satisfy
Eq. (\ref{cons1}). In Fig. 2 (b), the curves for $|r|$ and $|\bar r|$ coincide
for all $\om$; this can be shown to hold if $\si =0$, no matter what the 
filling fractions and velocities are. Figs. 2 (c) and (d) show the cases of 
$\si \ne 0$ for equal filling fractions and different filling fractions 
respectively; we have assumed for simplicity that $\si$ itself does not 
depend on $\om$. In this case, $k$ is complex as shown in Eq. (\ref{kom}) and 
(\ref{KA3}); the imaginary part of $k$ remains finite for large $\om$. Hence 
the different amplitudes show oscillations but they also decay as $\om$ 
increases. In the $\om \to 0$ limit, $r$ is given by Eq. (\ref{r1}).

Note that we have treated the fully interacting problem in this section, 
but in the DC limit $\om \to 0$, we recover the results for the 
non-interacting case given in Ref. \cite{sen}. This is because in that
limit, the terms involving $\pa_t \rho$ vanish in Eq. (\ref{eqm2}). The
currents $J_i (x)$ can then be found from Eq. (\ref{eqm2}) alone, and Eq.
(\ref{eqm1}) becomes unnecessary. The values of the currents therefore
do not depend on the interaction parameter $\la$ and the velocities $v_i$
appearing in Eq. (\ref{eqm1}). However, for the AC case, Eq. (\ref{eqm1})
and (\ref{eqm2}) are both required to find the currents, and the results 
are different for the interacting and non-interacting cases in general. It 
is also interesting to note that if there is no tunneling ($\si = 0$), the 
corrections to lowest order in the AC frequency for the reflection and
transmission amplitudes in Eq. (\ref{trom}) are of order $w$, but if there is 
tunneling ($\si \ne 0$), the lowest order corrections are of order 
$\om^{1/2}$. This follows from Eq. (\ref{kom}) which shows that for small 
$\om$, $k \sim \om$ if $\si = 0$, but $k \sim \om^{1/2}$ if $\si \ne 0$.

\begin{figure}[t] \begin{center}
\includegraphics[width=1.0 \linewidth]{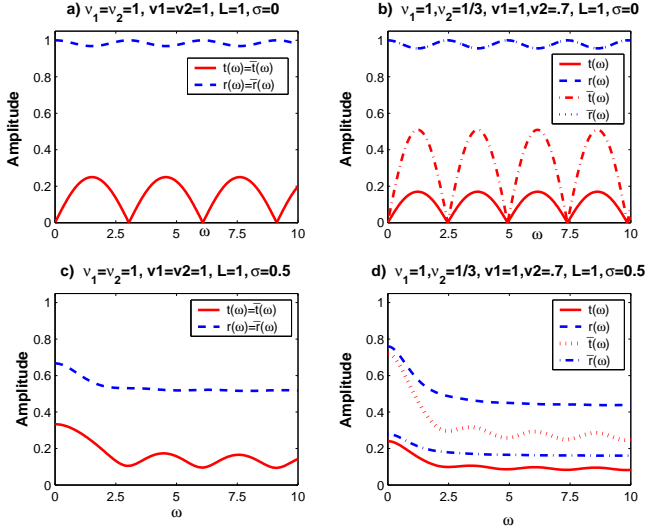} \end{center}
\caption{(Color online) Absolute values of AC current splitting amplitudes as 
functions of frequency $\om$ for a counter propagating line junction, for 
interaction $\la = 0.5$. Figures (a) and (b) correspond to tunneling $\si = 
0$, while (c) and (d) are for $\si = 0.5$. The units of $\om$, $\la$ and $\si$
are explained in the text.} \end{figure}

\section{Co-propagating case}

In the absence of density-density interactions between the co-propagating
modes, the Lagrangian is given by 
\bea {\cal L} &=& \frac{1}{4\pi \nu_1} ~\int_0^L dx ~\pa_x \phi_1~ (- \pa_t -
v_1 \pa_x) ~\phi_1 \non \\
& & + ~\frac{1}{4\pi \nu_2} ~\int_0^L dx ~\pa_x \phi_2 ~(-\pa_t - v_2 \pa_x) ~
\phi_2. \label{lag4} \eea
Here both the edge modes are taken to be propagating from $x=0$ to $L$. The 
corresponding density and current fields are defined as $\rho_{1/2}= \pa_x 
\phi_{1/2}/(2\pi)$ and $J_{1/2}= ~ -\pa_t \phi_{1/2}/(2\pi)$. The short-range 
repulsive density-density interaction between the two edges takes the form 
\bea {\cal L}_{int} &=& - ~\frac{\la}{4\pi \sqrt{\nu_1 \nu_2}} ~\int_0^L 
dx ~\pa_x \phi_1~ \pa_x \phi_2, \label{lag5} \eea
where $\la$ is the interaction strength (positive for repulsive interactions) 
with the dimensions of velocity. 

The equations of motion for the Lagrangian corresponding to Eq. (\ref{lag4}) 
and (\ref{lag5}), written in terms of the density and the current fields, are 
\bea J_1 ~-~ v_1 \rho_1 ~-~ \frac{\lambda \nu_1}{2 \sqrt{\nu_1 \nu_2}}~
\rho_2 &=& 0, \non \\
J_2 ~-~ v_2 \rho_2 ~-~ \frac{\lambda \nu_2}{2 \sqrt{\nu_1 \nu_2}}~ \rho_1 &=&
0. \label{eqm4} \eea

The tunneling along the LJ will be modeled using the following equations
\bea \pa_t \rho_1 ~+~ \pa_x J_1 &=& \frac{\si h}{e^2}\left(\frac{J_2}{\nu_2}
-\frac{J_1}{\nu_1} \right), \non \\ 
\pa_t \rho_2 ~+~ \pa_x J_2 &=& - \frac{\si h}{e^2}\left(\frac{J_2}{\nu_2}
-\frac{J_1}{\nu_1} \right). \label{eqm5} \eea

We can combine Eq. (\ref{eqm4}) and (\ref{eqm5}) to give 
\bea \label{eqm6}
\left(\pa_t+v_1 \pa_x + \frac{\al}{\nu_1}\right) J_1 &+& \left( \frac{\la 
\nu_1}{2 \sqrt{\nu_1\nu_2}} \pa_x - \frac{\al}{\nu_2} \right) J_2 ~=~0, \non \\
\left( \pa_t +v_2 \pa_x + \frac{\beta}{\nu_2} \right) J_2 &+&
\left( \frac{\la \nu_2}{2 \sqrt{\nu_1\nu_2}} \pa_x - \frac{\beta}{\nu_1}
\right) J_1 ~=~ 0, \non \\
& & \eea
where 
\bea \al &=& \frac{\si h}{e^2} ~(v_1-\frac{\la \nu_1}{2 \sqrt{\nu_1\nu_2}}), 
\non \\
\beta &=& \frac{\si h}{e^2} ~(v_2-\frac{\la \nu_2}{2 \sqrt{\nu_1\nu_2}}). \eea
Solving these equations with the appropriate boundary conditions gives us the 
current splitting matrix.

For the DC case, in a non-interacting system, the current splitting matrix
is given by Eq. (\ref{sdc}), and $\ga$ is given by \cite{sen}
\beq \ga = \frac{1 - e^{-L/l_c}}{2} ~~~ {\rm and} ~~~ l_c^{-1} = \frac{\si h}{
e^2} \left( \frac{1}{\nu_1}+\frac{1}{\nu_2} \right). \label{tdc2} \eeq

We now turn to the AC case. For the simplest case of same filling fraction 
($\nu_1=\nu_2=\nu$), same velocity ($v_1=v_2=v$)and and no tunneling ($\si=0$),
we obtain
\bea t(\om) =& \frac{2 \left(e^{i L k_1}-e^{i L k_2} \right) \left(\om - v 
k_1\right) \left(\om -v k_2\right)}{\lambda \om \left(k_1-k_2 \right)}, \non \\
r(\om) =& \frac{\left(\om -v k_2\right) k_1 e^{-i L k_1} - \left(\om -v k_1
\right)k_2 e^{-i L k_2} }{\om \left(k_1-k_2\right)}, \eea
where $k_1=\om/(v-\la/2)$ and $k_2=\om/(v+\la/2)$. Note that this is the 
dissipationless case and the amplitudes satisfy Eq. (\ref{cons1}), and in the
DC limit, $r \to 1$. 

For the case of same filing fractions ($\nu_1=\nu_2=\nu$), same velocities
($v_1=v_2=v$), but $\si\neq 0$, we get
\bea t(\om)&=& [2 i(e^{i L k_1}-e^{iLk_2})(\al -i \nu (\om -v k_1))(\al -i\nu
\non \\ 
&\quad& (\om -v k_2)) ] / [(2 v \al \nu +\lambda \nu(\al -i \nu \om ))(k_1
-k_2)] \non \\ 
r(\om) &=& [(e^{i L k_2}-e^{iLk_1})\left( 2\al (\al -i\nu\om ) + \la v \nu^2 
k_1 k_2 \right) \non \\ 
&\quad& + i \la (\al-i\nu\om)(k_1 e^{i L k_1}-k_2 e^{i L k_2})+ 2 i v \al \nu
\non \\ 
&\quad& (k_1 e^{i L k_2}-k_2 e^{i L k_1})]/ [ i \nu (2 \al v +\lambda (\al -
i \nu \om )) \non \\ 
&\quad& (k_1-k_2)]l, \eea 
where the values of $k_{1/2}$ are given by Eq. (\ref{KA7}).

\begin{figure}[t] \begin{center}
\includegraphics[width=1.0 \linewidth]{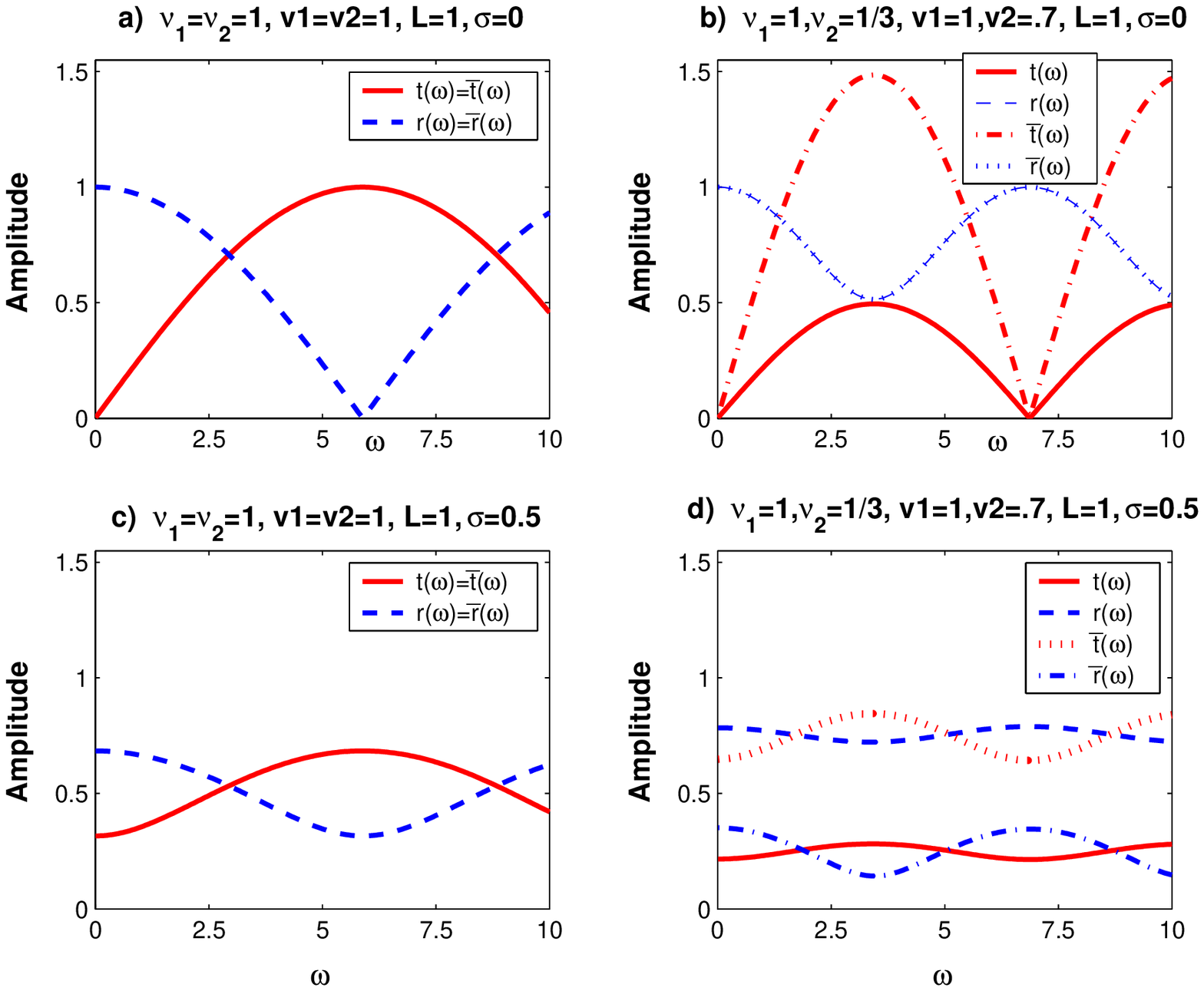} \end{center}
\caption{(Color online) Absolute values of AC current splitting amplitudes as 
functions of frequency $\om$ for a co-propagating line junction, for 
interaction $\la = 0.5$. Figures (a) and (b) correspond to tunneling $\si = 
0$, while (c) and (d) are for $\si = 0.5$. The units of $\om$, $\la$ and $\si$
are explained in the text.} \end{figure}

The expression for the most general case is given in the Appendix in Eq.
(\ref{KA7}) and (\ref{TRA9}). In the DC limit $\om \to 0$, Eq. (\ref{KA7})
gives $k_1 \to il_c^{-1} = i(\si h /e^2) (\nu_1^{-1} + \nu_2^{-1})$ and 
$k_2 \to 0$, while Eq. (\ref{TRA9}) gives 
\beq r(\om \to 0) ~=~ \frac{\nu_1+\nu_2~e^{- L/ l_c}}{\nu_1+\nu_2}. \eeq
Comparing with the expression in Eq. (\ref{sdc}), we find the same value of
$\ga$ as in Eq. (\ref{tdc2}).

In Fig. 3, we show the absolute values of the various reflection and 
transmission amplitudes as functions of the frequency $\om$ (in units
of $v_1/L = 1$) for various choices of $\nu_i$, $v_i$, $L$, $\la$ (in units 
of $v_1 = 1$), and $\si$ (in units of $e^2/(hL)$). Figs. 3 (a) 
and (b) show the case of $\si = 0$ (zero tunneling) for equal filling 
fractions and different filling fractions respectively, while Figs. 2 (c) and 
(d) show the case of $\si \ne 0$, again assuming that $\si$ does not
depend on $\om$. As in Fig. 2, we see prominent oscillations 
as a function of $\om$ when $\si = 0$, and both oscillations and decay when 
$\si \ne 0$. Just as in Fig. 2 (b), we note that the curves for $|r|$ and 
$|\bar r|$ coincide for all $\om$ in Fig. 3 (b) also.

Once again we note that we have treated the fully interacting problem here, 
but in the DC limit we recover the non-interacting results given in Ref. 
\cite{sen}. The reasons for this are the same as those explained at the
end of Sec. III. For the AC case, the expression for the current splitting 
matrix is different for the interacting and the non-interacting cases.

\section{Renormalization group and experimental implications}

Before discussing how measurements of the AC reflection and transmission
amplitudes can provide information about the parameters of the system such
as the tunneling conductance $\si$, the interaction strength $\la$ and the
edge mode velocities $v_i$, we have to consider the renormalization group (RG)
flow of $\si$. For the DC case, this has been discussed in Ref. \cite{sen}.
Briefly, the tunneling amplitude is given by a term in the Hamiltonian density
\beq {\cal H}_{tun} ~=~ \xi (x) ~\psi_1^\dag (x) \psi_2 (x) ~+~ h.c.,
\label{htun} \eeq
where $\psi_i (x)$ denotes the electron annihilation operator at point $x$ on
edge $i$ of the LJ. The tunneling conductance $\si$ is proportional to 
$|\xi|^2$ when $|\xi|$ is small. The presence of impurities near the LJ makes 
$\xi (x)$ a random complex variable; let us assume it to be a Gaussian 
variable with a variance $W$. Then $W$ satisfies an RG equation; to lowest 
order (i.e., for small $\xi$), this is given by \cite{giamarchi2,kane1}
\beq \frac{dW}{d \ln l} ~=~ (3 - 2d_t) ~W, \label{rg} \eeq
where $l$ denotes the length scale, and $d_t$ is the scaling dimension of the 
tunneling operator $\psi_1^\dag \psi_2$ appearing in Eq. (\ref{htun}). We 
will present expressions for $d_t$ below for both counter and co-propagating 
cases. There is also an RG equation for the interaction strength $\la$, but 
that can be ignored if $W$ is small.

Next, let us assume that the phase de-coherence length $L_T = \hbar v/(k_B T)$
(the length beyond which electrons lose phase coherence due to thermal
smearing) is much smaller than both the length $L$ and the scattering mean 
free path $L_m$ of the LJ. Successive backscattering events are then 
incoherent, and quantum interference effects of disorder are absent. One can 
then show \cite{kane1} that $\si$ scales with the temperature $T$ as 
$T^{2d_t - 2}$. It therefore seems that $\si L \to 0$ as $T \to 0$ if
$d_t > 1$. However, it turns out that this is true only if $d_t > 3/2$, i.e.,
if $W$ is an irrelevant variable according to Eq. (\ref{rg}). If $d_t > 3/2$
(called the metallic phase), one can simultaneously have $L \gg L_T$ (which 
justifies cutting off the RG flow at $L_T$ rather than at $L$), and $\si L 
\to 0$, i.e., $LT \gg 1$ and $L T^{2d_t - 2} \to 0$, for some range of 
temperatures. Further, $L_m$ scales with temperature \cite{kane1} as 
$T^{2-2d_t}$ and $L_T \sim T^{-1}$. Thus throughout the metallic phase, $L_m 
\gg L_T$ as $T \to 0$. But if $d_t < 3/2$ (i.e., $W$ is a relevant variable), 
we have $L/L_T \sim LT \gg 1$ and $T^{2d_t - 3} \to \infty$; hence $\si L 
\sim L T^{2d_t - 2} \to \infty$ (we call this the insulating phase).

The above analysis breaks down if one goes to very low temperatures where $L_T
\gtrsim L$ or $L_m$. In that case, the RG flow of $W$ has to be cut off at the
length scale $L$ or $L_m$, rather than $L_T$; hence $\si$ and therefore the
scattering coefficient $\ga$ become independent of the temperature $T$.

The scaling dimension $d_t$ can be computed using bosonization \cite{sen}.
For the counter propagating case, we find that 
\bea d_t &=& \frac{1}{4K} \left[ (1+K^2) \left( \frac{1}{\nu_1} +
\frac{1}{\nu_2} \right) - \frac{2(1-K^2)}{\sqrt{\nu_1 \nu_2}} \right], \non \\
{\rm where} ~~K &=& \sqrt{\frac{v_1+v_2-\la}{v_1+v_2+\la}}. \label{dtk} \eea
Thus $d_t$ depends on the interaction strength $\la$ and the velocities $v_i$.
(The stability of the system requires that $4 v_1 v_2 > \la^2$). For the 
co-propagating case, we have
\beq d_t ~=~ \frac{1}{2 \nu_1} ~+~ \frac{1}{2 \nu_2}, \eeq
which is independent of $\la$ and the $v_i$.

We can now discuss the experimental implications of our results for the 
various reflection and transmission amplitudes as a function of the
temperature. As mentioned earlier, a gate voltage can be used to control the 
distance between the two edges of the LJ. Making the gate voltage less 
repulsive for electrons is expected to reduce the distance between the edges;
this should increase both the strength of the density-density interactions as 
well as the tunneling conductance \cite{kane1}. In this way, one may be able 
to vary the scaling dimension $d_t$ across the value $3/2$ for the case of 
counter propagating edges in the LJ. We then see that quite different things 
should occur depending on whether the system is in the metallic phase ($d_t 
> 3/2$) or in the insulating phase ($d_t < 3/2$). In the metallic phase, $\si 
L \to 0$ as $T \to 0$; based on Figs. 2 and 3, we expect that oscillations 
versus $\om$ of the various amplitudes should become more prominent at
low temperatures. In the insulating phases, $\si L \to \infty$ as $T \to 0$;
Figs. 2 and 3 then indicate that the oscillations versus $\om$ of the 
different amplitudes should become less prominent at low temperatures. It
would be interesting to check this qualitative prediction experimentally.

\section{Discussion}

In this paper, we have discussed the response of a LJ separating two QH 
states to an AC voltage in the linear response (or small bias) regime. 
Depending on the filling fractions on the two sides of the LJ, the edges 
of the LJ may be counter propagating or co-propagating. We have presented 
a microscopic model for the system which includes short-range density-density 
interactions and electron tunneling between the two edges. The AC response 
can be described by a current splitting matrix $S_{ac}$; we have presented 
expressions for this matrix in terms of the AC frequency $\om$, 
the length of the LJ, the tunneling conductance, the strength of the 
interaction between the edges of the LJ, the filling fractions, and the 
velocities of the modes on the two sides of the LJ. In general, the elements 
of $S_{ac}$ oscillate with the frequency $\om$; the amplitude of oscillations
depends on $\om$ and the tunneling conductance $\si$ across the LJ. We 
find the interesting result that the matrix $S_{ac}$ does not depend on the 
interaction strength and the velocities in the DC limit $\om \to 0$, but 
does depend on those parameters for non-zero frequencies. [The fact that
the DC conductivity is independent of the interaction strength has been
observed earlier in the context of quantum wires modeled as non-chiral
Tomonaga-Luttinger liquids 
\cite{safi2,safi3,maslov,ponomarenko,safi1,safi4,furusaki,rech}.]
The low temperature behaviors of the elements of $S_{ac}$ can then be 
predicted based on a renormalization group analysis. In the case of counter
propagating edges, we find that depending on the interaction strength, the
system can be in either a metallic phase or an insulating phase. The two 
phases exhibit quite different behaviors of $S_{ac}$ as we go to low 
temperatures.

We emphasize that in the absence of any tunneling between the two edges, 
our calculation is valid in the linear response (small AC amplitude) regime 
and only for frequencies $\om$ which lie within the linearization 
regime for each LL wire (i.e., $\om < v/\al$). 
In the presence of inter-edge tunneling also, we
have assumed that the current is proportional to the potential (voltage) 
at every point with no phase difference between the two (Eq. (\ref{iv})). 
This is only true if $\om$ is less than the inverse of the relaxation time 
$\tau$ (for equilibration after each tunneling event). 
However, in the absence of a detailed theory of equilibration at finite 
frequencies, we do not know the precise form of $\tau$. 
Another limitation of our calculation is that the intrinsic frequency 
dependence of $\si$ is not known, although the RG analysis discussed in 
Sec. V gives an idea of the length scale dependence of $\si$ arising due 
to the interactions. We note that our various expressions for $t(\om)$ 
and $r(\om)$ will remain valid even if we take $\si$ to be frequency 
dependent. However, Figs. 2 (c-d) and 3 (c-d) have been made under the 
assumption that $\si$ does not depend on $\om$.

Before ending, we would like to mention some other studies of QH systems
with multiple filling fractions \cite{sandler,lal,rosenow1}. It may be 
interesting to extend our studies of AC response to these systems. Finally,
we note that the AC response of non-chiral Tomonaga-Luttinger liquids with
disorder has been studied earlier in some papers 
\cite{maura,fogler,rosenow2,safibena}.

\section*{Acknowledgments}

A.A. thanks Yuval Oreg for suggesting this problem and CSIR, India for 
financial support. We thank Sumathi Rao for comments and DST, India for 
financial support under Project No. SR/S2/CMP-27/2006.

\appendix
\section{Details of calculations}

We first find the current splitting matrix for the counter propagating case.
We start with the following guess for the edge currents along the LJ,
\bea J_1 &=& (a_1 e^{i k_1 x} + a_2 e^{i k_2 x}) ~e^{-i \om t}, \non \\
J_2 &=& (b_1 e^{i k_1 x} + b_2 e^{i k_2 x}) ~e^{-i \om t}. \label{sol1} \eea
These guess solutions must satisfy Eq. (\ref{eqm3}) for all values of $x$ 
along the LJ. This gives us the following equations
\bea &(-i \om + i v_1 k_1 + \frac{\al}{\nu_1})a_1+(\frac{i k_1 \la \nu_1}{2
\sqrt{\nu_1 \nu_2}} -\frac{\al}{\nu_2})b_1=0 , \non \\
&(-i \om + i v_1 k_2 + \frac{\al}{\nu_1})a_2+(\frac{i k_2 \la \nu_1}{2
\sqrt{\nu_1 \nu_2}} -\frac{\al}{\nu_2})b_2=0, \eea
where $\al=(\si h/e^2)(v_1+\la \nu_1/(2 \sqrt{\nu_1 \nu_2}))$ and $k_{1/2}$ 
is given by 
\bea \label{KA3} 
k_{1/2} &=& \frac{1}{2 \tilde{v}^2}[\om (v_2-v_1)+ i \tilde{v}^2 l_c^{-1} \pm 
[ (\om (v_2-v_1)+ i \tilde{v}^2 l_c^{-1})^2 \non \\ 
&+& 4 \tilde{v}^2 \left(\om^2+\frac{i\si h \om}{e^2} \left(\frac{v_1}{\nu_1}
+\frac{v_2}{\nu_2}+\frac{\la}{\sqrt{\nu_1 \nu_2}}\right)\right)]^{1/2}],
\eea
where $\tilde{v}=\sqrt{v_1 v_2 - \la^2/4}$, and $l_c^{-1}= (\si h /e^2)
(\nu_1^{-1} - \nu_2^{-2})$. Now consider an incoming current incident on the 
LJ from edge 1, and no incoming current from edge 3. Then we have the 
following equations at the two end points the LJ corresponding to the matrix 
$S_0$ defined in Eq. (\ref{s01}),
\bea a_1+a_2 &=& 1,~~~~~~ a_1 e^{ik_1 L} + a_2 e^{i k_2 l} = r(\om), \non \\
b_1+b_2 &=& t(\om),~~~ b_1 e^{ik_1 L} + b_2 e^{i k_2 l} = 0. \eea
Solving these six simultaneous equations gives us two elements of the AC 
current splitting matrix. For the most general case, we obtain
\bea \label{TRA5}
t(\om) &=& [2 \nu_2 (-e^{i L k_1}+e^{i L k_2}) (-i \om +i k_1 v_1+ \al/\nu_1)
(-i\om \non \\
&+& i k_2 v_1+ \al/\nu_1)]/ [e^{i L k_2} (-i \om +i k_2 v_1 + \al/\nu_1)
(2 \al \non \\
&-& i \lambda k_1 \sqrt{\nu_1 \nu_2}) - e^{i L k_1} (-i \om +i k_1 v_1 + \al/
\nu_1) (2 \al \non \\
&-& i \lambda k_2 \sqrt{\nu_1 \nu_2})], \non \\ 
r(\om) &=& [e^{i L k_1} (k_1-k_2) \nu_1 (\lambda \nu_2 (i \al +\om \nu_1) +2 
i \al v_1 \sqrt{\nu_1 \nu_2})] \non \\ 
&/& [2 (-1+e^{i L (k_1-k_2)}) \al (\al -i \om \nu_1) \sqrt{\nu_1 \nu_2} - k_2
\nu_1 (\lambda \nu_2 \non \\
&\times& (i \al +\om \nu_1)e^{i L (k_1-k_2)} +2 i \al v_1 \sqrt{\nu_1 \nu_2})
+k_1 \nu_1 (\lambda \nu_2 \non \\ & \times& (i \al + \om \nu_1 +k_2 v_1 \nu_1
(-1+e^{i L (k_1-k_2)}) ) +2 i\al v_1 \non \\
&\times& \sqrt{\nu_1 \nu_2} e^{i L (k_1-k_2)} )]. \non \\ \eea
Repeating this calculation for the case with an incoming unit current in wire 
3 and no incoming current in wire 1 will give us the other two amplitudes, 
$\bar{t}(\om)$ and $\bar{r} (\om)$, of the AC current splitting matrix.

For the co-propagating case, we again start from the guess solution given by 
Eq. (\ref{sol1}). Substituting them in Eq. (\ref{eqm4}), we get the following 
equations
\bea &(-i \om + i v_1 k_1 + \frac{\al}{\nu_1})a_1+(\frac{i k_1 \la \nu_1}{2
\sqrt{\nu_1 \nu_2}} -\frac{\al}{\nu_2}) b_1=0, \non \\
&(-i \om + i v_1 k_2 + \frac{\al}{\nu_1})a_2+(\frac{i k_2 \la \nu_1}{2
\sqrt{\nu_1 \nu_2}} -\frac{\al}{\nu_2}) b_2=0, \eea
where $\al=(\si h/e^2)(v_1-\la \nu_1/(2 \sqrt{\nu_1 \nu_2}))$ and $k_{1/2}$ 
is given by 
\bea \label{KA7}
k_{1/2} &=& \frac{1}{2 \tilde{v}^2}[\om (v_2+v_1)+ i \tilde{v}^2 l_c^{-1} \pm 
[ (\om (v_2+v_1)+ i \tilde{v}^2 l_c^{-1})^2 \non \\ 
&-& 4 \tilde{v}^2 \left(\om^2+\frac{i\si h \om}{e^2}\left(\frac{v_1}{\nu_1}+
\frac{v_2}{\nu_2}-\frac{\la}{\sqrt{\nu_1 \nu_2}}\right)\right)]^{1/2}]. \eea
where $\tilde{v}=\sqrt{v_1 v_2 - \la^2/4}$, and $l_c^{-1}= (\si h /e^2)
(\nu_1^{-1}+\nu_2^{-2})$. From the boundary conditions, we get
\bea a_1+a_2 =1,~& a_1 e^{ik_1 L} + a_2 e^{i k_2 l} =r(\om), \non \\
b_1+b_2 =0,~& b_1 e^{ik_1 L} + b_2 e^{i k_2 l} =t(\om). \eea
Solving these six simultaneous equations gives us 
\bea \label{TRA9}
t(\om) &=& [2 i\nu_2 (e^{i L k_1}-e^{i L k_2}) (\al -i \nu_1(\om -k_1 v_1))
(\al -i\nu_1 \non \\ 
&~\times& (\om -k_2 v_1))]/[(k_1-k_2) \nu_1 (2 \al v_1 \nu_1+\lambda 
\sqrt{\nu_1 \nu _2} \non \\ 
&~\times& (\al -i \om \nu_1) )], \non \\ 
r(\om) &=& [2 i\al \sqrt{\nu_1 \nu_2}(e^{i L k_1}-e^{iLk_2}) (\al -i\om \nu_1)
+k_1 \nu_1 (\lambda \nu_2 \non \\ 
& ~\times & (-i \nu_1 k_2 v_1 e^{i L k_2}+ (\al -i\om\nu_1 +i \nu_1 k_2 v_1)
e^{i L k_1} ) \non \\ 
&~+& 2 \al v_1 \sqrt{\nu_1 \nu_2} e^{i L k_2})-k_2 \nu_1 (\lambda \nu_2 (\al 
-i \om \nu_1)e^{iL k_2} \non \\ 
&~+& 2 \al v_1 \sqrt{\nu_1 \nu_2} e^{i L k_1})]/[(k_1-k_2) \nu_1 (\lambda 
\nu_2 (\al -i\om \nu_1)\non \\
&~+& 2 \al v_1 \sqrt{\nu_1 \nu_2} )]. \eea

\end{document}